\documentclass[10pt]{article}
\usepackage[cp1251]{inputenc}
\usepackage{graphicx}
\usepackage[russian]{babel}

\begin{document}

\twocolumn[\noindent{\small\it  ISSN 1063-7729, Astronomy Reports,
Vol. 52, No. 1, 2008, pp. 12--18. \copyright Pleiades Publishing,
Ltd., 2008. \noindent Original Russian Text \copyright A.T.
Bajkova, 2008, published in Astronomicheski$\check{\imath}$
Zhurnal, 2008, Vol. 85, No. 1, pp. 15--22. }

\vskip -4mm

\begin{tabular}{llllllllllllllllllllllllllllllllllllllllllllllll}
 & & & & & & & & & & & & & & & & & & & & & & & & & & & & & & & & & & & & & & & \\
\hline \hline
\end{tabular}

\vskip 1.5cm

\centerline{\Large\bf Structure of the Radio Source 3C 120 at 8.4
GHz from VLBA+ }\centerline{\Large\bf Observations in 2002 }

\bigskip

\centerline{\bf\large  A. T. Bajkova$^1$ and A. B.
Pushkarev$^{1,2}$}

\medskip

\centerline{\it $^1$Main Astronomical Observatory, Russian Academy
of Sciences, St.Petersburg, 196140 Russia}

\centerline{\it $^2$Crimean Astrophysical Observatory, Ukrainian
Academy of Sciences, Nauchnyi, Crimea, 98409 Ukraine}

\centerline{\small Received March 5, 2007; in final form, June 22,
2007}

\vskip 0.5cm

{\bf Abstract} --- { Abstract—Maps of the radio source 3C 120
obtained from VLBA+ observations at 8.4 GHz at five epochs in
January–September 2002 are presented. The images were
reconstructed using the maximum entropy method and the Pulkovo
VLBImager software package for VLBI mapping. Apparent superluminal
motions of the brightest jet knots have been estimated. The speeds
of jet knots decreases with distance from the core, changing from
$5.40 \pm 0.48c $ to $2.00 \pm 0.48c$ over 10 mas (where c is the
speed of light) for a Hubble constant of 65 km s$^{-1}$Mpc$^{-1}$.
This can be explained by interaction of the jet with the medium
through which it propagates.}

\bigskip


{\bf DOI}: 10.1134/S1063772908010022

\vskip 1cm

]

\centerline{1.~INTRODUCTION }

\medskip

The hypothesis that the jets in quasars and active galactic nuclei
are associated with the energy released by the accretion of matter
onto a central black hole via an accretion disk is strongly
supported by observations of microquasars in our own Galaxy [1].
Radio observations of microquasars display apparent superluminal
motions of bright jet components that appear after a sharp dip in
the X-ray radiation, believed to be due to a loss of the accretion
stability.

The first extragalactic source in which such processes were
detected was the active radio galaxy 3C 120 (redshift z = 0.033),
whose X-ray and radio emission displays behavior similar to that
observed in microquasars. Marscher et al. [2] showed that, after
X-ray dips, superluminal ejections propagate in the radio jets of
3C 120 with apparent speeds from 4.1c to 5.0c, where c is the
speed of light, for a standard cosmological model with a Hubble
constant of 65 km s$^{-1}$ Mpc$^{-1}$. In this case, the distance
to 3C 120 is 140 Mpc, and 1 mas corresponds to a projected
distance of 0.70 pc. The study [2] was carried out using VLBA+
observations at 43 GHz from November 1997 to April 2001 (16 epochs
in total).

In the current paper, we analyze the structure of the radio galaxy
3C 120 based on VLBA+ observations for five epochs in
January–September 2002 at 8.4 GHz and estimate the apparent speeds
of the brightest knots in the jet. The mapping employed the
Maximum Entropy Method (MEM), enabling us to obtain superresolved
images, and thereby trace the motion of the brightest knots in the
jet with higher accuracy.

\vskip 1cm

\centerline{2.~DATA}

\medskip

General information on the radio galaxy 3C 120 is given in Table
1. The observational data we used, obtained on the VLBA+ network,
were retrieved from the NRAO archive (USA). Information on these
observational data, obtained at five epochs from January to
September 2002, is given in Table 2. The VLBA+ network includes
the 10 stations of the Very Long Baseline Array (VLBA) together
with several stations of the global VLBI network. The stations in
Table 2 are given as two-character codes, whose full names can be
found at the site http://www
/evlbi.org/proposals/2\_lett\_station\_codes.txt. The filling of
the $UV$ (spatial frequency, or baseline) plane and the visibility
amplitude as a function of the baseline length projected onto the
UV plane are given in Fig. 1.

\vskip 1cm

\centerline{3. DATA-PROCESSING TECHNIQUES}

\medskip

The data were processed using the Pulkovo software package for
VLBI mapping VLBImager, which was developed by one of the authors
of this paper. The images were constructed using self-calibration
[3] in combination with MEM as the deconvolution procedure [4, 5].
Our choice of MEM as the deconvolution algorithm is based on our
desire to obtain superresolution under the condition that the
solution display the maximum smoothness. This makes it possible to
study finer structure in the source, and to determine the
coordinates of jet knots to higher accuracy [3]. We also used the
well-known technique of Gaussian model fitting [3] to find the
positions of jet knots. The errors in the coordinates were
estimated using the Monte-Carlo method, by varying the source
spectra within preset limits determined by the errors in the
visibility function [3].

\vskip 1cm

\centerline{4. RESULTS}

\medskip

Figure 2 presents maps of 3C 120 obtained by convolving the MEM
solutions with the “clean” beams indicated in the lower left
corner of each map. The contour levels on all maps in this paper
are 0.5, 1, 2, 4, 8, 16, 32, 64, and 90 \% of the peak value. The
parameters of the maps are listed in Table 3. Figures 3a and 3b
show the variations with epoch of the peak and total flux
densities in the radio maps. Note that the maps in Fig. 2 are in
good consistency with maps obtained from the same data using
various software packages (DIFMAP, AIPS; see the site
http://rorf.usno.navy.mil/RRFID/).

However, we are more interested in the direct MEM solutions, which
have a considerably higher resolution, and so display finer
features in the maps, while simultaneously being maximally smooth,
in accordance with the criterion of the maximum-entropy method.

Figure 4 presents the MEM images of 3C 120, arranged from top to
bottom in time order; the separations between the maps are
proportional to the time differences between the epochs. The
positions of four distinct knots in the jet (labelled A1, A2, A3,
and A4) are shown. Straight lines drawn from top to bottom and
connecting corresponding knots at different epochs show their
trajectories.

Table 4 lists the distances of these jet knots from the core,
labelled C in Fig. 4. The upper limit of the error in the
distances estimated via the Monte-Carlo method is approximately
$\pm 0.1$ mas, and we will use this as the maximum uncertainty in
the individual distances of the jet knots. The obtained
coordinates for components A1, A2, A3, and A4 as functions of the
epoch can be used to estimate their apparent speeds. The proper
motions in mas/year and corresponding speeds in units of the speed
of light for a standard cosmological model with cold dark matter
($\Lambda$ CDM) and a Hubble constant $H = 65$ km s$^{-1}$
Mpc$^{-1}$ are presented in Table 5. Figure 5 shows a graphical
representation of the motions of the components. Their apparent
speeds as functions of distance from the core are shown in Fig. 6.
Our derived speeds for components A1 and A2 (the closest to the
core) are consistent with the results of [2], while the motions of
the more distant components A3 and A4 are slower.

Thus, Fig. 6 demonstrates that the jet components do not move at
constant speeds: the closer the component to the core, the higher
its speed. On scales out to 10 mas (corresponding to 7 pc for the
adopted cosmology), the component speeds vary from $(5.40 \pm
0.48)c$ to $(2.00 \pm 0.48)c$. We took into account the error
corridor of $\pm 0.1$ mas for the straight lines shown in Fig. 5
when estimating the uncertainties of these superluminal speeds.
The deceleration in the component speeds with distance from the
radio core can be explained by interaction of the jet with the
medium through which it propagates: the deceleration will be more
noticeable the lower the emission frequency (the greater the
distance from the core). Therefore, the distance dependence of the
jet component speeds is much less pronounced in the 43-GHz maps
[2].

\vskip 1cm

\centerline{5. CONCLUSION}

\medskip

Our study of the structure of the radio galaxy 3C 120 in the
period from January to September 2002 based on VLBA+ observations
at 8.4 GHz has allowed us to determine apparent superluminal
speeds for the brightest knots in the extended jet on scales out
to 7 pc for a standard cosmological model with cold dark matter
and a Hubble constant of $H = 65$ km s$^{-1}$ Mpc$^{-1}$. We find
that the apparent speeds of the jet components are not constant,
and depend strongly on the distance from the radio core: the speed
decreases with increasing distance from the core. The apparent
speeds of the brightest jet knots decrease from $(5.40 ± 0.48)c$
to $(2.00 ± 0.48)c$, where c is the speed of light. This
deceleration can be explained by interaction of the jet with the
medium through which it propagates.

\vskip 1cm

\centerline{REFERENCES }

\medskip

\noindent 1.~I. F. Mirabel and L. E. Rodriguez, Nature {\bf 392},
673 (1998).

\vskip 1mm

\noindent 2.~A. P.Marsher et al., Nature {\bf 417}, 623 (2002).

\vskip 1mm

\noindent 3.~{\it Synthesis Imaging in Radio Astronomy II}, Ed. by
G. B. Taylor, C. L. Carilli, and R. A. Perley, Astron. Soc. Pac.
Conf. Ser. {\bf 180} (1999).

\vskip 1mm

\noindent 4.~A. T. Bajkova, Soobshch. Inst. Prikl. Astron., {\bf
58} (1993).

\vskip 1mm

\noindent 5.~A. T. Bajkova, Astron. Zh. {\bf 82}, 1087 (2005)
[Astron. Rep. {\bf 49}, 947 (2005)].

\vskip 2mm

{\noindent \it Translated by G. Rudnitskii }

\onecolumn

\newpage

\centerline{\bf Table 1. General information about 3C 120}
\begin{center}
\begin{tabular}{|c|c|c|c|c|c|}\hline

        & Right&  & Total flux&  & Redshift\\
Alias   & ascension& Declination & density& Optical & \\

   & (J2000)& (J2000) & at 5 GHz, Jy& counterpart & \\

\hline J0433+0521 &    4h 33m 11.0955s & +5$^{\circ}$ 21$^{'}$
15.62$^"$ & 3.8  & galaxy & 0.033\\
\hline
\end{tabular}
\end{center}

\vskip 0.5cm

\centerline{\bf Table 2. The VLBA+ observations of 3C 120}
\begin{center}
\begin{tabular}{|c|c|c|c|c|c|c|}\hline

 No. &Date & Frequency,& Polari- & Number & Station & Number\\
    &Epoch& MHz &zation  & of stations &names &of measurements\\\hline

 1.& 16.01.2002& 8409.97&  RCP & 18 & BR,FD,GC,HN,KP,LA,MC,& 4287\\
   & 2002.044 &         &        &    & MK,NL,NY,ON,OV,PT,SC,&  \\
    & && & &  TS,WF,KK,WZ& \\\hline
 2.& 06.03.2002&8409.97 & RCP & 18 & BR,FD,GC,HN,KP,LA,MC,&4784\\
   & 2002.178  &     & & &  MK,NL,NT,ON,OV,PT,SC,& \\
   & & & & &            TS,WF,KK,WZ& \\\hline
 3.&08.05.2002& 8409.97& RCP &17 & AR,BR,FD,GC,HH,HN,KP,& 5682\\
   &2002.351  & & & &  LA,MA,MC, NL,OV,PT,SC,& \\
   & & & & &      WF,KK,WZ & \\\hline
 4.&24.07.2002& 8409.97& RCP &15&BR,GC,HN,KP,LA,MA,MK,& 16556\\
   &2002.562  & & & &  NL,NY,OV,PT,SC,TC,WF,KK & \\\hline
 5.&25.09.2002 & 8409.97 &RCP& 15 & BR,FD,GC,KK,KP,LA,MA, &1186 \\
   &2002.734 & & & &  MC,MK,NL,ON,OV,PT,TS,WZ &\\\hline
\end{tabular}
\end{center}

\vskip 0.5cm

\centerline{\bf Table 3. Map parameters as functions of the
epoch}
\begin{center}
\begin{tabular}{|c|c|c|c|}\hline
      & Clean beam size &  Peak & Total \\
Epoch & (FWHM), & flux density, & flux density, \\
      & (mas$\times$ mas) & Jy/beam    &   Jy  \\\hline
2002.044  &  1.71$\times$0.63,-5.63$^\circ$ & 0.48 & 1.88 \\
2002.178 & 1.93$\times$0.66,-1.88$^\circ$ & 0.64 & 2.39 \\
2002.351 & 1.75$\times$0.94,10.80$^\circ$ & 1.09 & 2.52 \\
2002.562 & 1.65$\times$0.82, 5.53$^\circ$ & 0.91 & 2.07 \\
2002.734 &1.50$\times$0.80, 0$^\circ$ & 0.79 & 1.71\\\hline
\end{tabular}
\end{center}

\vskip 0.5cm

\centerline{\bf Table 4. Distance r from the core to the jet knots
A1–A4 }
\begin{center}
\begin{tabular}{|c|c|c|c|c|}\hline

Эпоха & A1 & A2 & A3 & A4 \\\hline

2002.044  & 1.79 &    4.22  & 5.91 & 8.67 \\\hline

2002.178  & 2.26  &   4.12 &  --     & 8.73\\\hline

2002.351  &   --    &  4.49 &  6.35  & 8.92\\\hline

2002.562  & 3.00   &  5.21 &--  & -- \\\hline

2002.734  & --      & 5.38  & -- & -- \\\hline
\end{tabular}
\end{center}

\vskip 0.5cm

\centerline{\bf Table 5. Apparent speeds of the jet knots}
\begin{center}
\begin{tabular}{|c|c|c|}\hline

Jet & Proper motion& Speed,\\
knot  &  mas/year&  c \\\hline A1 & 2.24$\pm0.20$ & 5.38$\pm0.48$
\\\hline A2 & 1.96$\pm0.20$ & 4.70$\pm0.48$ \\\hline A3 & 1.43$\pm0.20$ & 3.43$\pm0.48$
\\\hline A4 & 0.83$\pm0.20$ & 1.99$\pm0.48$
\\\hline
\end{tabular}
\end{center}

\newpage
\begin{figure}[t]
{\begin{center}
 \includegraphics[width= 90mm]{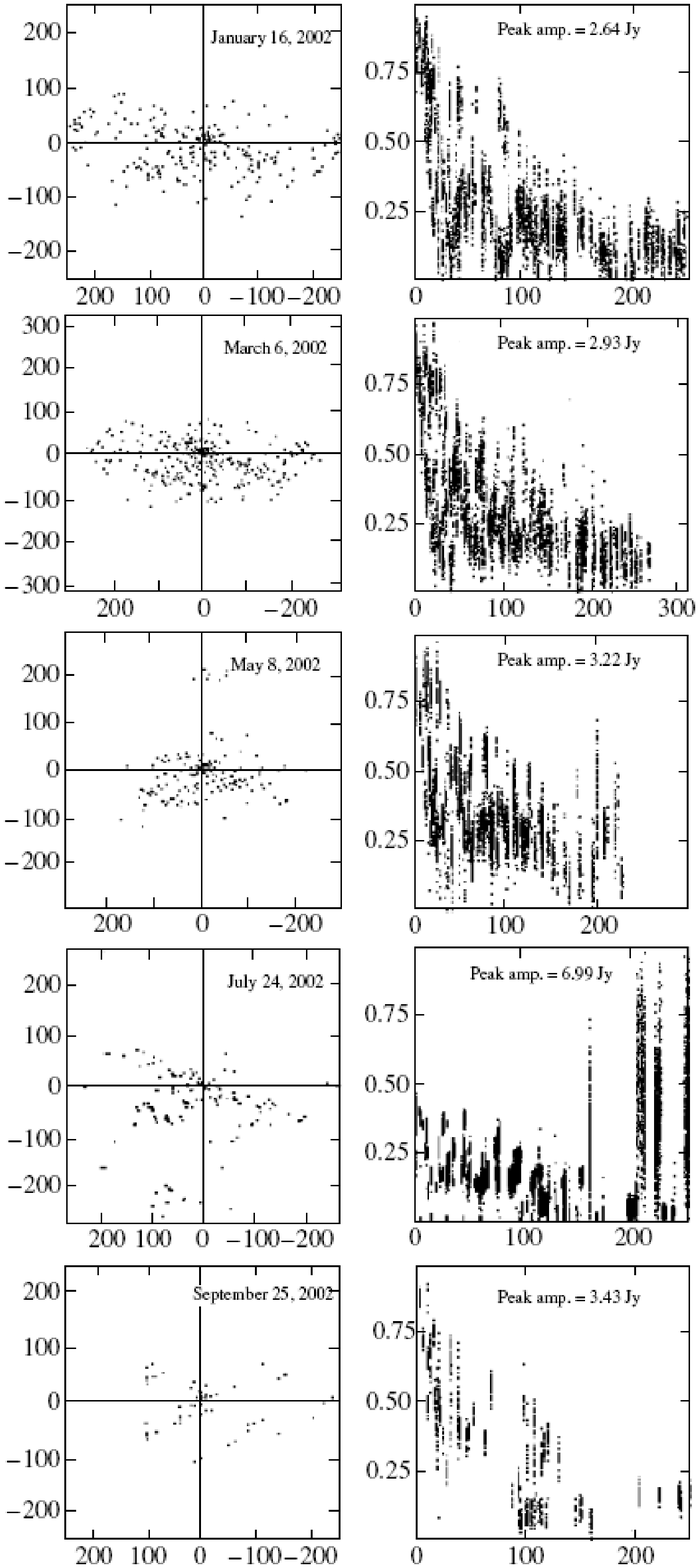}
\end{center}}
\begin{center} {\bf Fig.~1.}
Left: Coverage of the UV (spatial-frequency, or baseline) plane.
The horizontal and vertical axes show the U and V baseline
components in units of 106 wavelengths. Right: Amplitude of the
visibility function (in relative units) as a function of the
projected baseline length in units of 106 wavelengths for the five
epochs of observations. The peak amplitudes in Jy are given on the
graphs.
\end{center}

\end{figure}

\newpage
\begin{figure}[t]
{\begin{center}
 \includegraphics[width= 80mm]{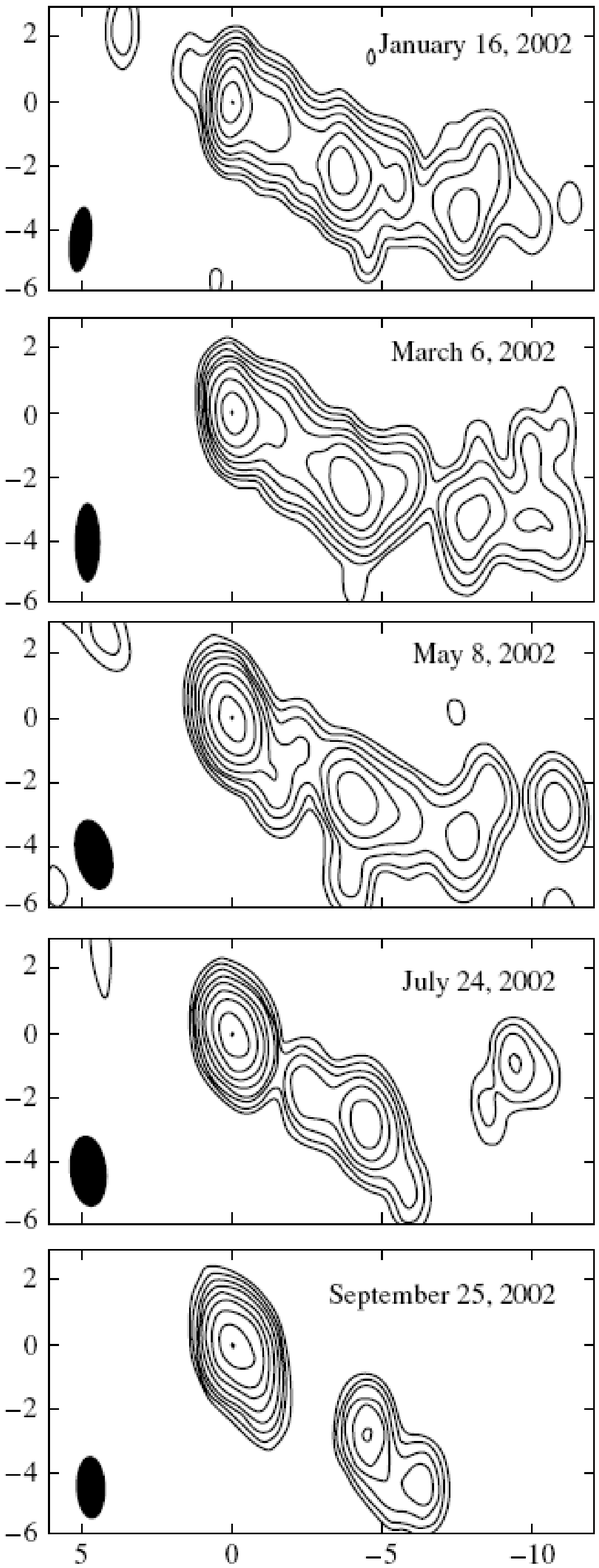}
\end{center}}
\begin{center} {\bf Fig.~2.}
Maps of 3C 120 obtained by smoothing the MEM solution using the
“clean” beams shown in the lower left corners of the maps. The
horizontal and vertical axes show relative right ascension and
declination in mas.
\end{center}
\end{figure}

\newpage
\begin{figure}[t]
{\begin{center}
 \includegraphics[width=100mm]{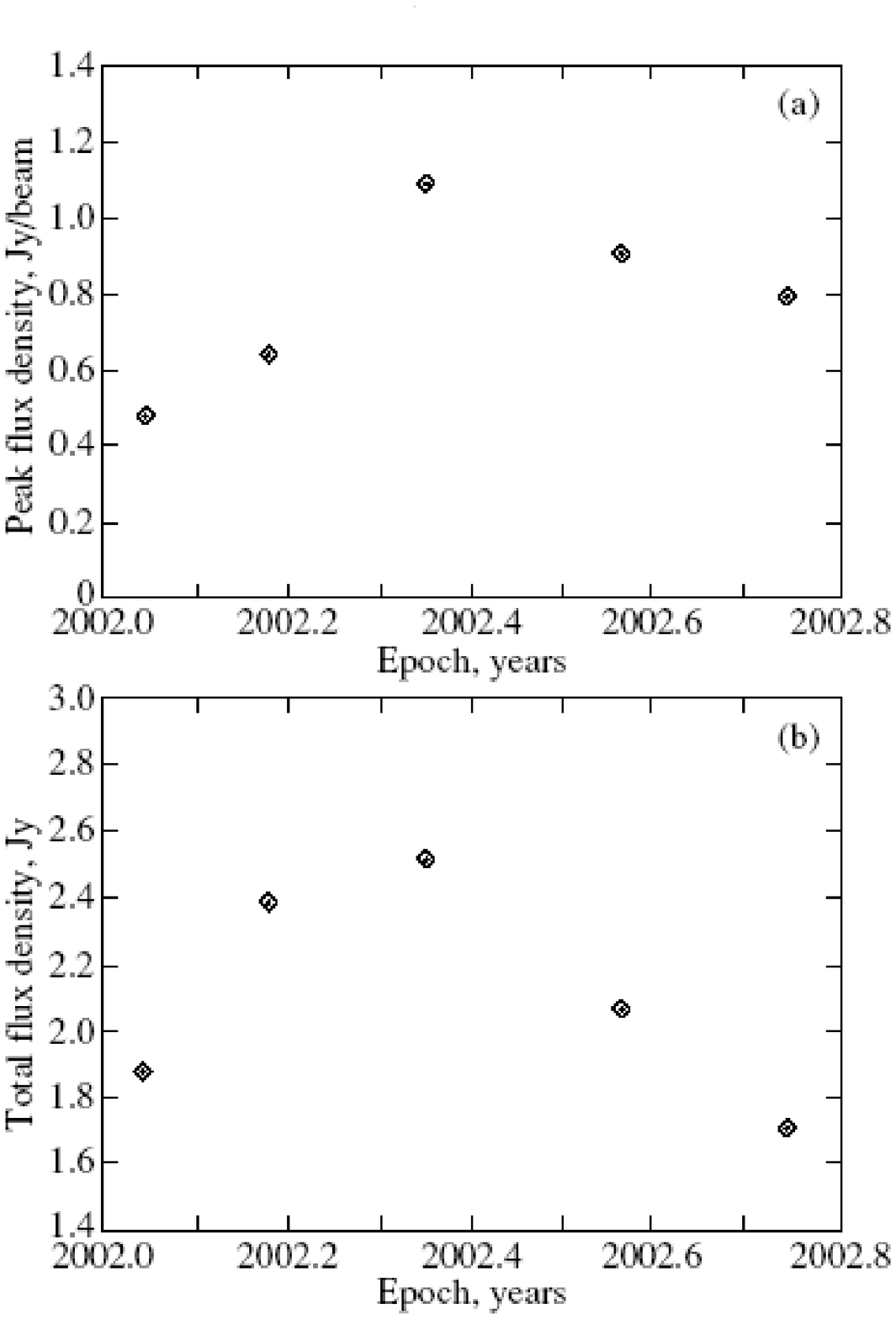}
\end{center}}

\begin{center} {\bf Fig.~3.}
(a) Peak and (b) total flux densities in the maps as a function of
the observing epoch.
\end{center}
\end{figure}

\newpage

\begin{figure}[t]
{\begin{center}
 \includegraphics[width=70mm]{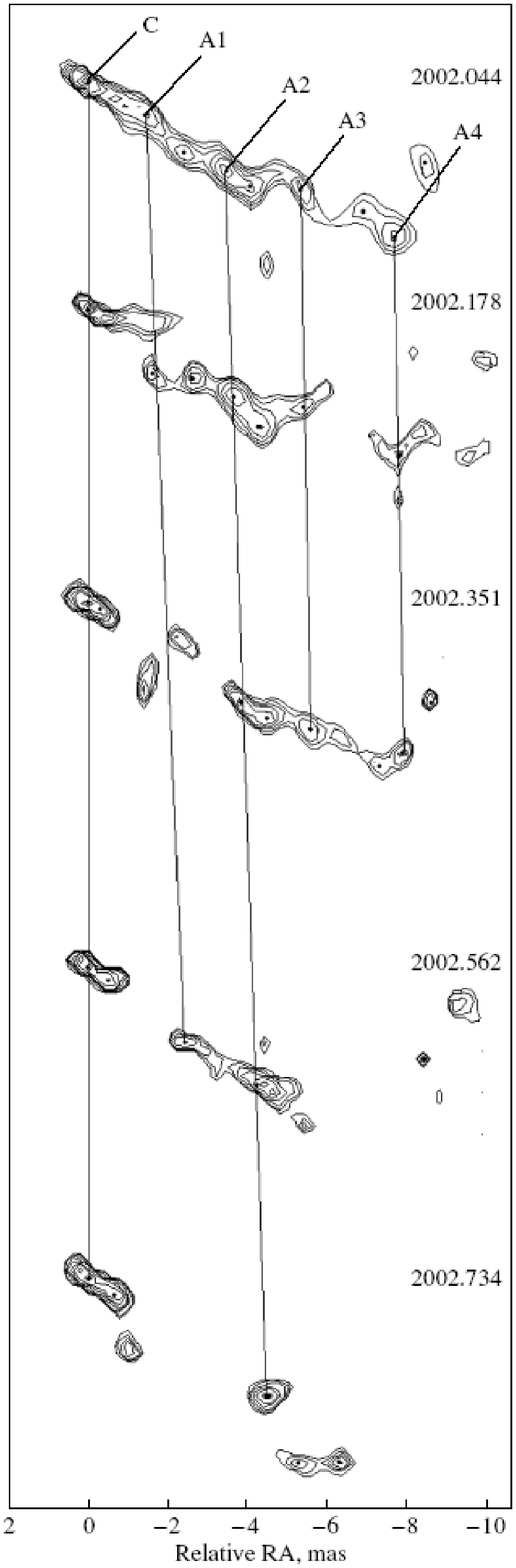}
\end{center}}
\begin{center} {\bf Fig.~4.}
Sequence of high-resolution MEM maps arranged from top to bottom
in time order. The distances between the map phase centers are
proportional to the time differences between the epochs. The
straight lines connecting the centers of the jet knots (labelled
A1–A4) show their trajectories.
\end{center}
\end{figure}

\newpage
\begin{figure}[t]
{\begin{center}
 \includegraphics[width=120mm]{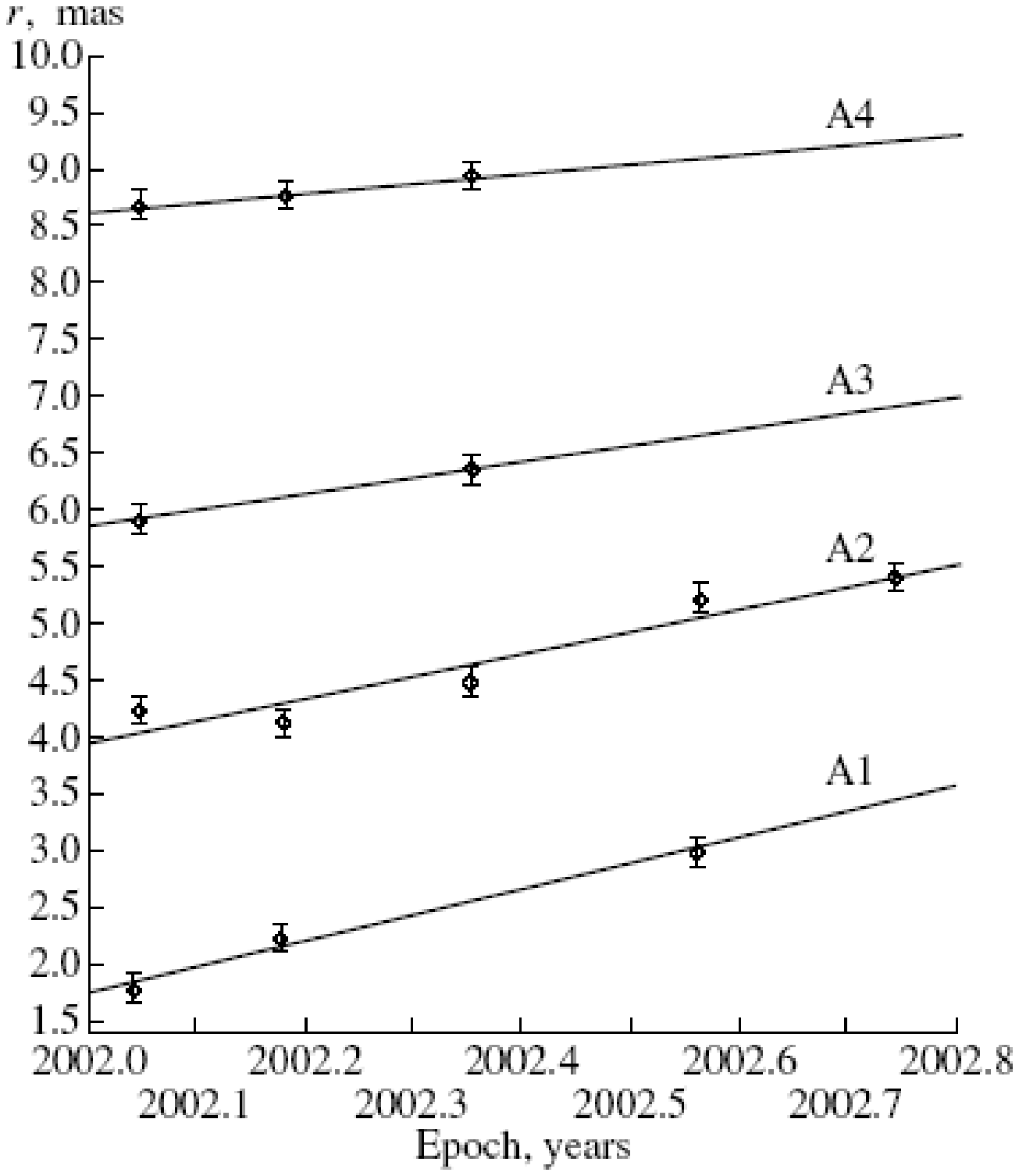}
\end{center}}
\begin{center} {\bf Fig.~5.}
Apparent motions of the jet knots A1–A4.
\end{center}
\end{figure}

\newpage
\begin{figure}[t]
{\begin{center}
 \includegraphics[width=120mm]{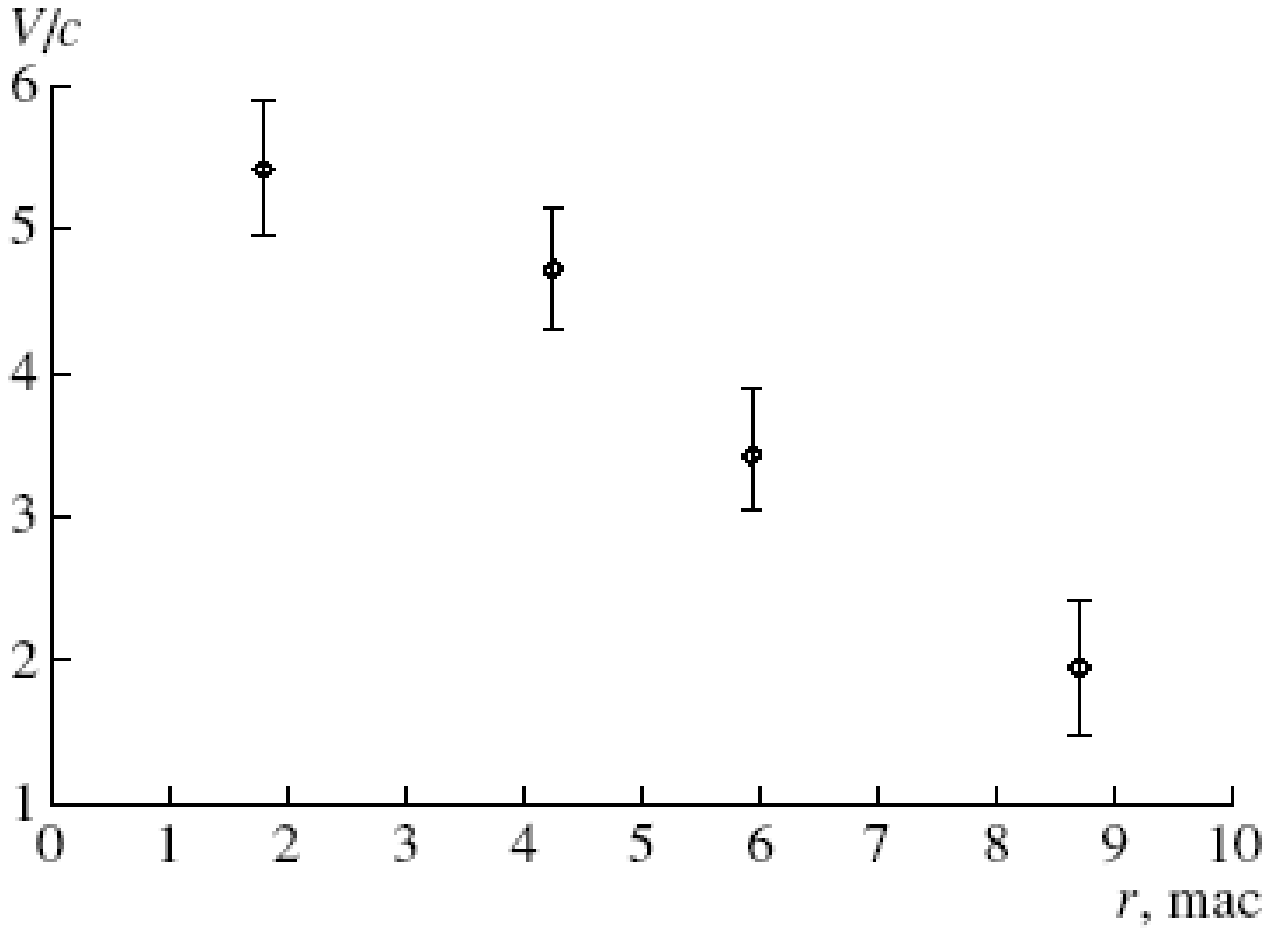}
\end{center}}
\begin{center} {\bf Fig.~6.}
Apparent speeds of the jet knots as a function of distance from
the radio core.
\end{center}
\end{figure}

\end{document}